# Study Overview for Super Proton-Proton Collider

Jingyu Tang[1], Yuhong Zhang[2], Qingjin Xu[1], Jie Gao[1], Xinchou Lou[1], Yifang Wang[1]

For the SPPC study group

[1] Institute of High Energy Physics, Chinese Academy of Sciences, Beijing 100049, China

[2] Thomas Jefferson National Accelerator Facility, Newport News, VA 23606, USA

**Abstract**: SPPC (Super Proton-Proton Collider) is a discovery machine that is designed for energy frontier research in two decades from now, as the second stage of the CEPC-SPPC project. The main objective is to carry out experiments at 125 TeV in center-of-mass energy in a two-ring collider of 100 km in circumference and 20 T in dipole field. This white paper about SPPC describes the machine related issues, including performance, design overview, design challenges, key technologies and their maturity and required R&D, staged options and upgrades, synergies with other facilities, and environmental impacts.

## 1. Design Overview

1.1 Introduction and status

SPPC (Super Proton-Proton Collider) is the second phase of the CEPC-SPPC project [1-2], with CEPC (Circular Electron-Positron Collider) as factory of Higgs boson and other particles for precision studies and SPPC as a discovery machine aiming at the energy frontier research beyond the Standard Model. Both machines will share a tunnel of 100 km in circumference. While CEPC is designed to have a center-of-mass (CM) energy of 60-350 GeV [3], SPPC can reach unprecedented high CM energy up to 125 TeV [4]. According to the presently envisioned timeline of the CEPC-SPPC project, CEPC will be constructed in 2028-2035 and starting of SPPC construction will likely be after 2044. In parallel to the CEPC design phases of PreCDR, CDR and the undergoing TDR, the study on SPPC has been also evolved and deepened. The current study on SPPC is equivalent to Pre-CDR.

1.2 Performance matrix

- **Energy**

Choice of SPPC CM energy is a compromise of several limitation factors including availability and down-selection of high field SC magnet technology, synchrotron radiation power, luminosity and budget of site power consumption. High-field magnets are the key technology for future hadron colliders. Presently SPPC has selected high-temperature superconductor (HTS) based magnet technology with a strong preference on iron-based superconductors for its great potential in attaining magnetic field of 20 T or even higher and cost effectiveness. The synchrotron radiation power on the beam screen which separates the cold magnet bore and the beam vacuum increase rapidly in a 4$^{th}$ power of the beam kinetic energy, and will cause various complex problems such as vacuum pumping, electron clouds, cryogenic heat load, impedance, and magnet quench protection. Suffering from lower circulation beam current or less accumulated protons to the limitation on synchrotron radiation power, the attainable integrated luminosity is lower at higher energy. Higher cryogenic heat load on the beam screen means higher power consumption. SPPC chooses 125 TeV in CM energy as its design goal utilizing 20-T magnets, and 75 TeV as the intermediate run stage with 12-T magnetic field to obtain higher luminosity.

- **Luminosity**

At tens of TeV energy, the proton beam emittance shrinks quite fast, in a decay time of about or less than 1 hour, leading to a rapid increase of the instant luminosity, which is usually beneficial to obtain higher integral luminosity. However, the peak luminosity is limited by the beam-beam effects and other beam physics issues, thus the emittance damping process should be controlled and mitigated by a heating mechanism.

Design of final focusing at IP is primarily limited by the long distance ($L^*$) between IP and the inner quadrupole triplet and performance of these final focusing quadrupoles. The former is due to large detector size, and the latter demands very high pole-tip fields and superior radiation shielding capability to mitigate severe radiation near IP. With a nominal $\beta^*$ of 0.5 m (0.75 m for the run at 75 TeV), dynamically decreasing $\beta^*$ to 0.25 m in the course of collision by taking advantage of emittance shrinking is also considered. A small crossing angle between the colliding beams at IP must be adopted to accommodate relatively small bunch spacing (20 ns) to prevent parasitic collisions at IP, however, it leads to a reduction of luminosity and also induces coupling between the transverse and longitudinal motions, thus crab cavities are needed to compensate the effects.

Sophisticated luminosity levelling and optimization will be needed to obtain high integral luminosity while limiting all the other detrimental effects. This advanced scheme involves a controlled emittance shrinking and performing dynamic $\beta^*$ squeeze, etc. With a circulation beam current of 0.19 A, the peak and annual integral luminosities are about $1.34\times10^{35}$ cm$^{-2}$s$^{-1}$ and 1.2 ab$^{-1}$, respectively, assuming 160 days run time per year and 70% machine availability.

- **Injector chain**

To feed two collider rings of 62.5 TeV in beam kinetic energy, a multi-stage injector complex is required and its conceptual design is currently in progress. Since the individual accelerators are all large-scale, it is envisioned that the construction of the SPPC injector complex will be started earlier than that of the collider rings to allow commissioning of the facility stage by stage. As shown schematically in Figure 1, the first stage of the injector, *p-Linac*, is a high power H-minus superconducting linac of 1.2 GeV in kinetic energy and 50 Hz in repetition rate. The second stage, *p-RCS*, is a rapid cycling synchrotron (RCS) of 10 GeV and 25 Hz. It is followed by the 3rd stage, *MSS* (Medium Stage Synchrotron) of 180 GeV and 0.5 Hz. The final stage of the injector, SS (Super Synchrotron), will boost the beam energy to 3.2 TeV, the injection energy of the SPPC collider rings. These accelerators all have high repetition rates and high average beam powers. SS also has high beam energy. They can be utilized to support additional physics programs using different beams after their commissioning. Even after the SPPC collider rings are constructed and pp collision begins, the multiple-beam programs can be continued since an injection of beam into a synchrotron takes relatively short time compared to its cycling period thanks to a higher repetition rate for the lower stage. In addition to accelerating beam to the injection energy of the collider rings, other key beam characteristics required by the collider are also prepared in the injector chain. The normalized transverse emittance of 1.2 μm.rad and bunch population of $4\times10^{10}$ are achieved by the phase space painting in p-RCS. Bunch spacing of 25 ns is achieved by the RF systems in p-RCS and MSS.

- **Facility scale and power consumption**

A number of candidate sites nationwide are currently under evaluation for hosting CEPC-SPPC. The comprehensive site evaluation includes geological condition, environmental and economic impact, availability of electric power supply, other utilities and transportation. A down-selection will be made likely at the time of the final Chinese government approval of the project. The SPPC collider rings share a deep underground tunnel of 100 km in circumference with CEPC. Depth of the tunnel is likely 50 to 100 m depending on the geological condition of the selected site. The SPPC injector, a four-accelerator complex, will be located inside the large collider rings, as illustrated in Figure 1. The dedicated tunnels for the injector accelerators with the longest one being about 8 km for the SS are near ground or in a more shallow level to save the construction cost and facilitate physics programs using external beams. There are also beam transport lines to connect the accelerators.

To support such a large accelerator complex and the experimental systems, conventional facilities are very important. Although CEPC will build the relevant conventional facilities to meet its needs, expanding the facility capacity will be necessary to meet operation of SPPC, even when we do not consider the coexistence of the two colliders. The total power consumption for SPPC is estimated on an order of 400 MW, about two thirds from the collider (accelerator and detectors) and the other one third from the injector complex. In the collider rings, the main contribution is from the cryogenic systems including those for the SC magnets, beam screens and SRF cavities. In the proton injector, the power supplies for the room-temperature magnets, cryogenic systems for the SC magnets in SS and the superconducting cavities in p-Linac, and the RF sources in all four accelerators have major contributions to power consumption. Green energy solutions will be in high priority starting from the CEPC.

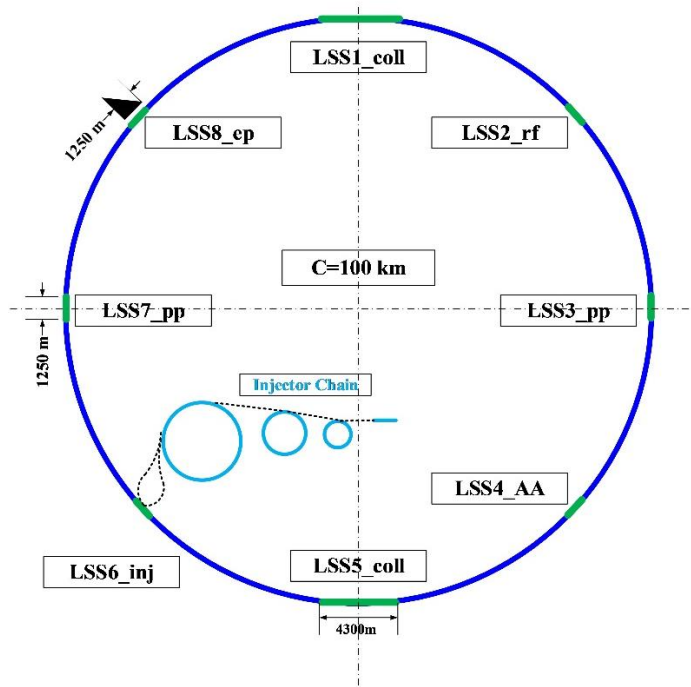

Fig. 1. Schematic layout of the SPPC complex. Dimension of the collider and three booster rings are in scale. Two interaction points are located at 3 and 9 clock.

## 1.3 Design Summary

Since the kick-off of the CEPC-SPPC study in 2012, SPPC has been positioned as the second stage of the project. The preliminary SPPC design study was included in two comprehensive design reports, namely, the CEPC-SPPC Pre-CDR and CDR. The latest SPPC design and recent R&D progress will be summarized in the upcoming CEPC-SPPC TDR scheduled to be released at the end of 2022 or the beginning of 2023. The SPPC baseline scheme has experienced the evolution of key design goal from 75 TeV in center-of-mass energy with a tunnel circumference of 54 km and a magnetic field of 20 T in the PreCDR, to 75 TeV with 100 km and 12 T in the CDR, to the current 125 TeV with 100 km and 20 T. From the study, a future proton-proton collider aiming for particle physics in energy frontier in two decades from now is outlined. Main accelerator physics problems and key technological challenges have been identified and studied, and key performance specifications including reach of the CM energy and annual integral luminosity have been studied. The SPPC main parameters are shown in Table 1.

Table 1: Main SPPC parameters

| Parameter | Value | Unit |
|---|---|---|
| **General design parameters** | | |
| Circumference | 100 | km |
| Beam energy | 62.5 | TeV |
| Lorentz gamma | 66631 | |

| Dipole field | 20 | T |
|---|---|---|
| Dipole curvature radius | 10415.4 | m |
| Arc filling factor | 0.78 | |
| Total dipole magnet length | 65.442 | km |
| Total arc length | 83.9 | km |
| Number of long straight sections | 8 | |
| Total length of straight sections | 16.1 | km |
| Energy gain factor in collider rings | 19.53 | |
| Injection energy | 3.2 | TeV |
| Number of IPs | 2 | |
| Revolution frequency | 3.00 | kHz |
| **Physics performance and beam parameters** | | |
| Initial luminosity per IP | $4.3 \times 10^{34}$ | $cm^{-2}s^{-1}$ |
| Beta function at collision points ($\beta^*$) | 0.5 | m |
| Circulating beam current | 0.19 | A |
| Nominal beam-beam tune shift limit per IP | 0.015 | |
| Bunch separation | 25 | ns |
| Number of bunches | 10080 | |
| Bunch population | $4 \times 10^{10}$ | |
| Accumulated particles per beam | $4 \times 10^{14}$ | |
| Normalized rms transverse emittance | 1.2 | μm |
| Beam life time due to burn-off | 8.1 | hours |
| Total inelastic cross section | 161 | mb |
| Full crab crossing angle | 73 | μrad |
| rms bunch length | 60 | mm |
| rms beam spot size at IP | 3.0 | μm |
| Stored energy per beam | 4.0 | GJ |
| SR power per beam | 2.2 | MW |
| SR heat load at arc per aperture | 26.3 | W/m |
| Energy loss per turn | 11.4 | MeV |

## 1.4 Design Challenges

- **Beam physics**

Although we can learn many things from the design and operation of LHC, a future proton-proton collider of several times larger in scale still faces many serious challenges in beam physics. Design of the final focus lattice is one of them. Due to the large detector size and very high radiation dose rate in the interaction region, it is required to design the lattice with careful considerations on the optimization of $\beta^*$, machine-detector interface (MDI), radiation protection, chromaticity correction, accommodation of crab cavities, etc. One also needs to consider the dynamic adjustment in $\beta^*$ or the crossing angle.

Beam collimation is another key issue, since the SPPC collimation system has to tackle unprecedentedly high stored energy in beam (a few GJ in each beam, about 10 GJ at maximum), which is a few tens times higher than that in LHC. A sophisticated collimation system with extremely high collimation efficiency must be developed in order to support safe operation of the machine.

In order to attain highest possible luminosity, the beam-beam parameters are pushed to the limit according to today's experience and understanding, efficient compensation or alleviation methods should be explored.

With an extremely high beam energy, especially after the beam energy is increased from the earlier 37.5 TeV to the current 62.5 TeV, synchrotron radiation in the SPPC collider rings becomes so important that it imposes critical technical problems to the vacuum system. One has to find an adequate beam screen solution that also causes serious beam physics problems such as impedance, electron cloud effects etc.

Impedance and instabilities are always important problems in larger colliders. At SPPC, it is a challenging task to control contribution of impedance in a very long ring circumference, in particular from a complicated collimator system and beam screen structures yet to be developed. Besides the measures in budgeting the global impedance, sophisticated beam control and manipulation including bunch-by-bunch feedback and emittance heating mechanism are needed.

There are strong concerns on beam injection to and ejection from the SPPC collider rings. The former is related to the bunch filling pattern and control of quality of the injected beam, and both are very important for the machine safety under abnormally functioned conditions.

Since all the injector synchrotrons have unprecedentedly high beam powers, there will be also obvious challenges in designing those accelerators, especially for very low loss control with beams of multiple megawatts.

- **Machine design**

The SPPC machine design must ensure compatibility with CEPC since the rings of both colliders share a common tunnel. If needed in the future, even the whole CEPC machine including the detectors should be kept during the construction of SPPC. This looks to be feasible with a complicated bypass design, but will certainly increase the difficulty of designing the SPPC.

As there is huge energy stored both in the beams and in the magnet systems, the abnormal release of energy poses critical challenges to safe operation of the machine. Even with the normal operation, managing the large beam loss power and ejection of the spent beams are critically important to hands-on maintenance and operation reliability.

Since the SPPC injector accelerators are so large and complex, it is very challenging to build and commission them almost simultaneously. Even the 100-km collider ring tunnel already exist during the SPPC phase, there will still be a large amount of civil constructions required for the SPPC injector complex and other supporting facilities. .

- **Required key technologies**

There will be many technological challenges in constructing such an accelerator complex. Since it is planned for the far future or two decades from now, one can imagine utilization of some advanced technologies that are beyond the state-of-the-art and might be developed from many years' R&D efforts. Here some of the key technologies are stressed.

➢ High-field magnets: dipole and quadrupole magnets with high field and high field quality are one of the critical challenges and the highest R&D priority for future energy frontier hadron colliders including SPPC. This is not only because their requirements are far beyond the present state-of-the-art, but also they are likely dominating cost drivers of the SPPC project. It is well understood that the total project cost will be a top deciding factor for a final approval of the project by the funding agency. Based on the preliminary investigations at IHEP, high-temperature superconductors (HTS), especially the type of iron-based superconducting (IBS), is considered to hold the best chance to reach unpresented 20-T magnetic field required by SPPC. In addition, this type of magnet technology could potentially be significantly more cost effective than the current $Nb_3Sn$ superconductors. Though the IBS technology currently is still in the early development phase, the SPPC team has already decided to focus the main R&D efforts in this direction.

➢ Beam screen and vacuum: high synchrotron radiation power from a proton beam of tens TeV energy induces critical problems to the vacuum system. As superconducting magnets are used, the advantage of using low-temperature vacuum tubes as cryogenic pumps suffers the heat load of very high level from synchrotron radiation in the arcs. In addition, the radiation also introduce serious electron cloud problem that may lead to beam instability. A beam screen is needed to shield the heat load to the vacuum tube that is at the same temperature level as the magnet cold bore or a few Kelvins. It is clear that the beam screen structure developed for LHC does not meet the requirements of SPPC, since the synchrotron radiation power in SPPC is about 100 times higher. It is a challenging task to develop an adequate beam screen structure that can solve the relevant problems such as vacuum pumping, electron cloud alleviation, low coupling impedance and mechanical sustaining of magnet quenches. In addition, if the magnets could work at a temperature higher than 2 K to take the advantage of HTS technology, it will be an important problem to know how to pump hydrogen molecules in the vacuum.

➢ Crab cavities must be used to recover the luminosity loss and mitigate the coupling effect between the transverse and longitudinal motions due to the crossing angle at IPs. High transverse kicking RF field and low impedance are two key design specifications of the crab cavities. In addition, these cavities must be highly compact since the space near IPs where the cavities will be installed is tight.

➢ Kickers with fast rise/fall times and high fields are required to inject the beams and eject the abort beams with relatively small time gaps between bunch trains. At the same time, superconducting septum magnets with radiation protection are also needed due to very high beam rigidity. All these special magnets requires new designs and R&D efforts.

➢ To tackle high beam loss power of MW level during normal operation and even more critical beam losses in the abnormal operation conditions, sophisticated beam collimation systems with extremely high collimation efficiency must be developed. To sustain the critical irradiation of the lost protons and secondary particles, the material of the collimators is of principal importance. In addition, since there are many collimators with some of them very close to the beams, impedance becomes major concern when designing the structure and choosing the material. Some R&D efforts are demanded in the future.

➢ Since many components or devices in the accelerator complex work at low temperature, such as superconducting magnets, superconducting RF cavities and beam screens, the majority of the power consumption is in the cryogenic systems. Increasing efficiency of the cryogenic plant and transmission of electric power is a key issue, which should be studied in the next two decades.

- **Environmental Impacts**

Civil construction: deep underground tunnels explosion and large amount of excavation will pose serious environment impact to the facility site and surrounding area. SPPC civil construction practice will be similar to CEPC, focusing on high efficiency and minimum environmental impact. New technologies will be aggressively pursued to achieve engineering and cost goals.

Radiation management: being a high energy proton machine, SPPC faces much high challenges than CEPC in radiation control and protection. Radiation shielding and management of radioactive waste are more demanding. A comprehensive plan for the facility-wide radiation management must be developed and properly implemented before commissioning and operating the facility.

Conventional facilities: this concerns about ground water, electricity, adaptation and protection of the ecological environment. This will be addressed in the period before the construction of CEPC, and there is not much change with adding the new machine – SPPC.

## 2. Technology Requirements

2.1 Technology Readiness Assessment

While most of accelerator and detector technologies required by hadron colliders already exist nowadays, many of them do need significant improvements, and some of them must be far beyond state-of-the-art, in order to meet the higher requirements of SPPC. Among many technical risk items, the highest one is availability of high-field (up to 20 T) superconducting magnets, which demands massive technological development.

## 2.2 Required R&D

As mentioned in Section 1.4, many key technologies for constructing the SPPC are beyond state-of-the-art and require R&D efforts. Some of them are within the scope of the global efforts outlined in the technical LOIs submitted to Snowmass 2021.

- **High-field magnets**

Long-term R&D efforts are needed to develop the superconducting materials and magnet technology of very high field. As the iron-based HTS technology that has great potential but is far from maturity is chosen for SPPC, both domestic and international collaborations on the R&D efforts, which are already underway, should be strengthened with a roadmap for phased development goals. Besides the superconducting cable, the magnet technology at 20 T despite of different kinds of superconductors does not exist today and also demands many years' development.

- **Beam screen and vacuum**

A beam screen structure that meets the multiple relevant requirements such as vacuum pumping and stability, electron cloud alleviation, low coupling impedance and mechanical sustaining of magnet quenches should developed. Besides the geometrical structure, the coating with special conductive material to the inner surface of the screen might be needed to reduce impedance when being resistant to the eddy current during the magnet quenches. The temperature for the screen is a dedicated issue concerning the cryogenic power consumption and vacuum instability.

- **Related R&D LOIs**

A number of LOIs (Letter of Interest) submitted to Snowmass21 are related to the R&D studies required for future proton-proton colliders including SPPC. They are listed here:

High-field SC magnets related: AF4-AF7-22 [5] and AF4-AF7-187 [6],

General technological requirements related: AF4_AF3-EF0-RF0-025 [7], AF7-AF4-64 [8]

Beam-material interaction related: AF7-AF4-159 [9]

MDI related: AF7-AF4-54 [10]

Heavy ion beam cooling related: AF7-AF4-108 [11]

## 2.3 Required and Desirable Demonstrators

No demonstrator is needed for designing and constructing the SPPC.

# 3. Staging options and upgrades

## 3.1 Energy upgrades

As it was outlined in the CEPC-SPPC CDR, SPPC will have two major phases. The first one has 75 TeV in CM energy by utilizing 12-T magnets. The second phase is the ultimate stage with a CM energy of 125-150 TeV by utilizing magnets of 20-24 T, depending on the outcome of the R&D program for achieving ultra-high magnet field strength. The current baseline scheme for SPPC is to design a collider with 125 TeV and 20 T. However, the intermediate run stage of 75 TeV can be kept just like the LHC Run-1 operating at 7 TeV.

Table 3: Comparison of the main parameters between the two energy runs

| Parameters | @75 TeV | @125 TeV | Unit |
|---|---|---|---|
| Dipole field | 12 | 20 | T |
| Initial luminosity per IP | $1.0 \times 10^{35}$ | $4.3 \times 10^{34}$ | $cm^{-2}s^{-1}$ |
| Annual integrated luminosity | 1.24 | 0.65 | $ab^{-1}$ |
| Circulating beam current | 0.73 | 0.19 | A |
| Bunch population | $1.5 \times 10^{11}$ | $4 \times 10^{10}$ | |
| Normalized emittance | 2.4 | 1.2 | mm-mrad |
| Stored energy per beam | 9.1 | 4.0 | GJ |
| SR power per beam | 1.1 | 2.2 | MW |
| SR heat load at arc per aperture | 12.8 | 26.3 | W/m |

## 3.2 Luminosity upgrades

Luminosity upgrades will be developed for the SPPC runs. Current study shows that higher luminosity is available at lower energy such as 75 TeV, which is especially beneficial for precision-related research. At 125 TeV, phased luminosity goals are also planned to allow the evolution of detector technology to tackle higher event rate and the machine-related methods to deal with the higher synchrotron radiation and instabilities, similar to the HL-LHC with respect to the LHC.

## 3.3 Experimental system upgrades

Detector technology is undergoing tremendous developments for different applications from high-energy physics to multidisciplinary research based on advanced light sources. It is foreseen that during the SPPC lifetime the detector systems will take major upgrading just like the CMS/ATLAS upgrade related with HL-LHC.

## 3.4 Possible electron-proton collider

*e-p* collisions can also be realized in the CEPC-SPPC complex by bringing one beam from each of two colliders together and converting two *pp* collision IR for *e-p* collisions. The CM energy of *e-p* collision could reach 6.7 TeV (by 62.5 TeV *p* × 180 GeV *e*), which is about 20 times higher than the CM energy of HERA, the world only ever-built *ep* collider, and nearly 50 times higher than the CM energy of EIC, a new polarized *ep/eA* collider currently under construction in US. Naturally, parameters of the CEPC-SPPC ep colliding beams are similar to these of *pp* and *e+e-* collisions, however, certain consideration must be taken in the conceptual design and modifications of several key parameters are necessary in order to meet the *e-p* physics requirements and facility restrictions. Three issues are briefly discussed below.

The first issue one must consider is total facility power. While both designs of CEPC and SPPC meet the administrative limit of individual total power consumption, running *e-p* collisions nevertheless requires operation of both facilities including their injector chains, sum of the powers required by all facilities clearly pose challenge in managing overall power budget. The good thing is that only one electron beam and one proton are needed. Thus, total lepton synchrotron radiation power is reduce by 50% and one of two proton rings which is made of high field SC magnets can be turned off. Both changes should enable substantial reduction of power consumption, thus possibly permits delivering sufficient currents of the electron and proton beam for *e-p* collisions. It is assumed at the present stage of the study the nominal currents of the electron beam in CEPC and of the proton in SPPC can be attained for *e-p* collisions and they are used for estimation of e-p collision luminosity performance. This assumption certainly should be validated by further study.

The second issues is selection of a common bunch frequency in two collider rings. While the SPPC rings store over ten thousand bunches, CEPC rings store a very small number of bunches at high energies due to very low beam current limited by strong synchrotron radiation, there are only 249 bunches in Higgs Factory run of CEPC as an example. This creates a huge mismatch of two colliding bunch trains. The preliminary study shows a medium bunch number (around 3000) should provide an optimized luminosity.

The third issue concerns crab crossing of electron and proton beams at an interaction point. CEPC utilizes a crab waist scheme for delivering ultra-high luminosity. Nevertheless, adopting this advanced IR scheme requires extremely flat colliding beams, this cannot be realized for the proton beam. Therefore, similar to the SPPC *pp* collisions, the scheme utilizing crab cavities is adopted to restore head-on collision in the CM frame in the *e-p* collision. In addition, since the proton beam is a round one, in order to better match the beam spots at collision points, the electron beam is made to a round one by introducing coupling of beam optics.

The following table summarizes the main parameters of CEPC-SPPC *e-p* collisions. Two design points are chosen, one is at the energy frontier (62.5 TeV *p* × 120 GeV *e*), the other illustrates high luminosity potential, up to $4.2 \times 10^{34}$ cm$^{-2}$s$^{-1}$ at one collision point, about 2800 times higher than the best luminosity HERA ever had achieved.

Table 4: CEPC-SPPC *e-p* main parameters

| Particle    |     | Proton | Electron | Proton | Electron |
|-------------|-----|--------|----------|--------|----------|
| Beam energy | TeV | 62.5   | 0.12     | 37.5   | 0.0455   |
| CM energy   | TeV | 5.48   |          | 2.61   |          |

| Beam current | mA | 165 | 16.7 | 310 | 803.5 |
|---|---|---|---|---|---|
| Particles per bunch | $10^{10}$ | 10.2 | 1.0 | 5.4 | 14 |
| Number of bunch | | 3360 | | 11951 | |
| Bunch spacing | Ns | 75 | 75 | 25 | 25 |
| Bunch repetition rate | MHz | 13.3 | 13.3 | 40 | 40 |
| Normalized emittance, (x/y) | μm rad | 1.2 | 150 | 2.4 | 80 |
| Bunch length, RMS | mm | 60 | 5 | 60 | 5 |
| Beta-star (x/y) | cm | 90 | 2.5 | 75 | 5 |
| Beam spot size at IP (x/y) | mm | 4.0 | 4.0 | 6.7 | 6.7 |
| Beam-beam parameter per IP (x/y) | | 0.0011 | 0.15 | 0.007 | 0.15 |
| Crossing angle | mrad | ~73 | | ~73 | |
| Hour-glass (HG) reduction factor | | 0.70 | | 0.87 | |
| Luminosity per IP, with HG factor | cm$^{-2}$s$^{-1}$ | $3.7 \times 10^{33}$ | | $4.2 \times 10^{34}$ | |

## 4. Synergies with other concepts and/or existing facilities

### 4.1 Synergies on machine technologies

Most knowledge about accelerator physics and technology for SPPC can be based on those developed during the design and operation of LHC. FCC-hh and SPPC were proposed almost simultaneously and two machines have comparable performance goals, there are many synergy aspects between them. High-field magnets and fast ramping superconducting magnets are demanded in different future projects. For example, SPPC, FCC-hh and the Muon Collider requires high field magnets for the collider rings, and fast ramping superconducting magnets for fast acceleration in the earlier stages of accelerator chains of the Muon Collider and to a less demand in SS of the SPPC. High power proton accelerators have been undergoing rapid developments over the last two decades and will continue in the next decade, which are related to spallation neutron sources, neutrino beams and ADS (Accelerator-Driven System) applications, and this will benefit the design and technological availability of the SPPC injector accelerators.

### 4.2 Synergies on detector technologies

The advance in detector technology is very important to attain high performance of experimental systems. The design and construction of the detectors at the SPPC will benefit from the development of the detectors for the colliders including CEPC, FCC-ee, ILC/CLIC and the Muon Collider and also non-collider detectors within the next two decades.

### 4.3 Synergies on conventional facilities and green power

Energy utilization efficiency becomes more and more impotent in construction and operation of large scientific facilities, and green energy technology is undergoing rapid progress. The CEPC and SPPC will benefit from the technological advances developed in other large scientific facilities. Technology advances in the areas of water purification and circulation, air conditioning and circulation, automatic installation and transport, etc. are also great interesting to the SPPC project.

## 4.4 Synergies for physics research

Although the SPPC has the highest center-of-mass energy among the existing proposals for the incoming decades, some of the other proposals also have the goal to carry out energy frontier research, in particular the mult-TeV colliders like FCC-hh, the Muon Collider and CLIC. In addition, SPPC can contribute to the Higgs physics research in higher orders, and this will be complementary to the proposed Higgs factories.